\newcommand{\bea}{\begin{eqnarray}}
\newcommand{\eea}{\end{eqnarray}}
\newcommand{\be}{\begin{equation}}
\newcommand{\ee}{\end{equation}}
\newcommand{\ben}{\begin{enumerate}}
\newcommand{\een}{\end{enumerate}}
\newcommand{\bi}{\begin{itemize}}
\newcommand{\ei}{\end{itemize}}
\newcommand{\bmi}[1]{\begin{minipage}{#1 cm}}
\newcommand{\emi}{\end{minipage}}

\def\elabel#1{\label{eq:#1}}

\def\eck#1{\left\lbrack #1 \right\rbrack}

\def\rund#1{\left( #1 \right)}
\def\abs#1{\left\vert #1 \right\vert}

\def\ave#1{\left\langle #1 \right\rangle}

\def\d{{\rm d}}

\def\vp{\varphi}
\def\vt{{\vartheta}}

\def\Real{{\rm I\mathchoice{\kern-0.70mm}{\kern-0.70mm}{\kern-0.65mm}%
  {\kern-0.50mm}R}}
\def\C{\rm C\kern-.42em\vrule width.03em height.58em depth-.02em
       \kern.4em}
\font \bolditalics = cmmib10
\def\bx#1{\leavevmode\thinspace\hbox{\vrule\vtop{\vbox{\hrule\kern1pt
        \hbox{\vphantom{\tt/}\thinspace{\bf#1}\thinspace}}
      \kern1pt\hrule}\vrule}\thinspace}

\def \vc #1{{\textfont1=\bolditalics \hbox{$\bf#1$}}}
{\catcode`\@=11
\gdef\SchlangeUnter#1#2{\lower2pt\vbox{\baselineskip 0pt \lineskip0pt
  \ialign{$\m@th#1\hfil##\hfil$\crcr#2\crcr\sim\crcr}}}
}

\def\ueber#1#2{{\setbox0=\hbox{$#1$}%
  \setbox1=\hbox to\wd0{\hss$\scriptscriptstyle #2$\hss}%
  \offinterlineskip
  \vbox{\box1\kern0.4mm\box0}}{}}

\def\bx#1{\leavevmode\thinspace\hbox{\vrule\vtop{\vbox{\hrule\kern1pt
        \hbox{\vphantom{\tt/}\thinspace{\bf#1}\thinspace}}
      \kern1pt\hrule}\vrule}\thinspace}

{\catcode`\@=11
\gdef\SchlangeUnter#1#2{\lower2pt\vbox{\baselineskip 0pt \lineskip0pt
  \ialign{$\m@th#1\hfil##\hfil$\crcr#2\crcr\sim\crcr}}}
}

\documentclass{aa}
\usepackage{graphicx}
\begin{document}
   \title{The consequences of parity symmetry for higher-order
   statistics of cosmic shear and other polar fields}

   \author{Peter Schneider
          \inst{1}
          }

   \offprints{P. Schneider}

   \institute{Institut f. Astrophysik u. Extr. Forschung, Universit\"at Bonn,
              Auf dem H\"ugel 71, D-53121 Bonn, Germany\\
              \email{peter@astro.uni-bonn.de}
             }

   \date{Received ; accepted }

   \abstract{We consider the parity transformation and the
consequences of parity invariance on the $n$-point correlation
function of the cosmic shear {caused by gravitational lensing by the
large-scale structure}, or any other polar (or `spin-2') field.
The decomposition of the shear field into E- and B-modes then yields
the result that any correlation function which contains an odd number
of B-mode shear components vanishes for parity-invariant random shear
fields. {In particular, this result implies that the expectation value
of the third-order cross aperture statistics, $\ave{M_\times^3}$,
vanishes for parity-invariant shear fields. Therefore, a significant
detection of a non-zero value of $\ave{M_\times^3}$ in a cosmic shear
survey does not indicate the presence of B-modes, but of an
underestimate of the statistical uncertainty or cosmic variance, or a
remaining systematic effect in the shear measurement.} We argue that
the parity invariance provides a very specific diagnostic of
systematic effects in shear data. {Our results apply as well to
the linear polarization of the cosmic microwave background.}
  \keywords{Dark matter -- Methods: statistical -- Cosmology:
miscellaneous -- cosmic microwave background}
}
   \titlerunning{Consequences of parity invariance of shear fields}
   \maketitle
%

\section{Introduction}
Cosmic shear, resulting from the gravitational deflection of light
bundles from distant sources by the tidal gravitational field of the
large-scale structure, has been recognized as an important tool for
observational cosmology (e.g., Blandford et al.\ 1991;
Miralda-Escud\'e 1991; Kaiser 1992; see Mellier 1999 and Bartelmann \&
Schneider 2001 for recent reviews). It allows the direct
investigation of the (dark) matter distribution up to redshifts $\sim
1$, without making assumptions about the relation between dark matter
and luminous tracers, like galaxies. Therefore, since the first detections
of cosmic shear (Bacon et al.\ 2000; Kaiser et al.\ 2000; van Waerbeke
et al.\ 2000; Wittman et al.\ 2000) intensive observational and
theoretical research has been performed.

Most of the work up to now has concentrated on second-order statistics
of the shear field. However, Bernardeau et al.\ (2002) have detected a
significant third-order shear signal in the Virmos-Descart cosmic
shear survey. It was realized before that the third-order cosmic shear
statistics would provide an invaluable tool for cosmological parameter
estimates; in particular, a third-order shear measurement breaks the
(near) degeneracy between the density parameter $\Omega_{\rm m}$ and
the power spectrum normalization $\sigma_8$ (Bernardeau et al.\ 1997;
Jain \& Seljak 1997; van Waerbeke et al.\ 1999). These theoretical
studies concentrated on the third-order statistics of the convergence
field, from which the shear field is derived. However, as the shear is
the observable (or more precisely, the observable galaxy ellipticities
provide an unbiased estimate of the shear), not the convergence,
recent interest has focused on determining the third-order statistical
properties of the shear itself. 

Since the shear is a two-component quantity, the three-point
correlation function (3PCF) has eight independent components and
depends on three arguments.  Bernardeau et al.\ (2003) considered a
specific combination of components of the 3PCF, calculated its
properties, both analytically and numerically, and applied this to
measure the 3PCF on the Virmos-Descart data (Bernardeau et al.\
2002). With the large number of
components of the 3PCF, it is not clear a priori which of them carry
most of the information about the underlying matter field. In contrast
to the two-point correlation function (2PCF), which has only two
non-vanishing components (the other two vanish by parity invariance),
none of the eight components of the 3PCF vanishes for a general set of
three points. Only for specific triangles (where two or three sides
are equal) does parity imply that some components vanish (Schneider \&
Lombardi 2003; hereafter SL03). This has later been verified using
ray-tracing simulations through a cosmological matter distribution, as
well as with analytic estimates of the 3PCF from the halo model
(Takada \& Jain 2003a; Zaldarriaga \& Scoccimarro 2003). As it turned
out, investigating the statistical properties of the 3PCF of a polar
field is far from trivial, and some confusion about symmetry
properties has arisen.

The purpose of this paper is to clarify some of the issues related to
parity transformations of the shear field. In Sect.\ts 2 we recall the
definition of the shear 3PCF and its behavior under parity
transformations. In Sect.\ts 3 the decomposition of a general shear
field into E- and B-modes (Crittenden et al.\ 2002; Schneider et al.\
2002) is introduced. The central result of this paper, derived in
Sect.\ts 4, is as follows: For a pure B-mode shear field whose
statistical properties are invariant under parity transformation
(hereafter: parity symmetric), the
3PCF vanishes identically. More generally, the $n$-point shear correlation
function containing $m$ B-mode shear components and $(n-m)$ E-mode
shear components vanishes identically for $m$ odd. {This result will be
shown both by an explicit calculation, as well as by a simple
heuristic, though accurate, consideration (at the beginning of Sect.\ts 4).}

\section{Parity transformation of the 3PCF}
\subsection{Definition of the 3PCF}
We consider a shear field (or more generally, any polar field) that is
statistically homogeneous and isotropic. This implies that the 3PCF
depends on three arguments only, which define the shape and size of a
triangle. Let $\vc X_i$, $i=\{1,2,3\}$, be three points defining a
triangle, and let $\vc x_1=\vc X_3-\vc X_2$, $\vc x_2=\vc X_1-\vc X_3$,
and $\vc x_3=\vc X_2-\vc X_1$ be the side vectors. Furthermore, let
$\gamma_\mu(\vc X_i)$, $\mu=\{1,2\}$, be the Cartesian components of
the shear at the point $\vc X_i$. We note that the Cartesian
components of the shear can also be
considered as components of a symmetric trace-free matrix,
\be
\Gamma=\rund{ \begin{array}{cc}
\gamma_1 & \gamma_2 \\
\gamma_2 & -\gamma_1 \\
\end{array} } \;,
\elabel{transf}
\ee
and its behaviour under linear coordinate transformations $\vc x\to
\vc x'= T \vc x$ is
\be
\Gamma'=T^{\rm T}\, \Gamma\, T\;.
\ee
Then, the Cartesian components of
the 3PCF are defined as 
\be
\gamma_{\mu\nu\lambda}(\vc x_1,\vc x_2,\vc x_3)\equiv
\ave{\gamma_\mu(\vc X_1)\,\gamma_\nu(\vc X_2)\,\gamma_\lambda(\vc
X_3)}\;,
\ee
where it was used that the shear field is a homogeneous random field,
so that the 3PCF is invariant under translations; therefore,
$\gamma_{\mu\nu\lambda}$ does depend only on the separation vectors
$\vc x_l$. In order to employ the assumed rotational invariance of the
shear field, one needs to measure the shear components relative to a
direction that rotates with the triangle. For example, one can
define the center of mass $\vc o=(\vc X_1+\vc X_2+\vc X_3)/3$ of the
triangle and define the shear components at $\vc X_i$ relative to the
direction vector $\vc X_i-\vc o$ between the point $\vc X_i$ and the
center $\vc o$. If $\zeta_i$ denotes the polar angle of the direction
vector $\vc X_i-\vc o$, we define the tangential and cross component
of the shear at $\vc X_i$ {\it relative to the center} $\vc o$ by
\bea
\gamma(\vc X_i,\zeta_i)&\equiv& \gamma_{\rm t}(\vc X_i,\zeta_i)+{\rm
i}\gamma_{\times}(\vc X_i,\zeta_i) :=-\gamma(\vc X_i) {\rm e}^{-2{\rm
i}\zeta_i} \nonumber \\
&=&-\eck{\gamma_1(\vc X_i)+{\rm i}\gamma_2(\vc X_i)}
{\rm e}^{-2{\rm i}\zeta_i}
\;,
\elabel{2}
\eea
as can be seen from (\ref{eq:transf}) when setting $T$ as a rotation
matrix.  We then define the components of the 3PCF relative to the
center of mass as
\be
\gamma^{\rm (cen)}_{\mu\nu\lambda}(x_1,x_2,x_3)=\ave{\gamma_\mu(\vc
X_1)\,\gamma_\nu(\vc X_2)\,\gamma_\lambda(\vc X_3)}\;,
\elabel{3}
\ee
where now $\mu,\nu,\lambda=\{{\rm t},\times\}$. It should be noted
that the 3PCF defined in this way is invariant under rotations of the
triangle, and thus depends only on the lengths $x_i=\abs{\vc x_i}$ of
the side vectors, provided that we fix the orientation such that $\vc
x_1 \times \vc x_2 = \vc x_2 \times \vc x_3 = \vc x_3 \times \vc x_1
>0$.

There is nothing special about the center of mass; instead, different
centers of the triangle (or other points attached to a triangle) can
be chosen in the definition of the tangential and cross components of
the shear, and thus in defining the 3PCF; SL03 considered some of the
most obvious choices (orthocenter, center of incircle, center of
circumcircle), and derived the relations between the 3PCF defined with
respect to different centers. 

\subsection{Parity transformation}
SL03 have derived the behavior of the 3PCF under parity
transformations. Consider a fixed triangle, and its mirror symmetric
sister obtained by flipping the original one along the line
perpendicular to the side vector $\vc x_3$ going through $\vc
X_3$. This transformation has two effects: first, the points $\vc X_1$ and
$\vc X_2$ are interchanged, so that the orientation of the triangle is
now negative, in the sense that $\vc x_1\times \vc x_2<0$. Second, the
tangential components of the shears are unaffected, whereas the cross
components change sign. Together, this implies for the behavior under
parity transformations that 
\be 
{\rm P}\eck{\gamma_{\mu\nu\lambda}^{\rm (cen)}(x_1,x_2,x_3)} =\Pi\,
\gamma_{\nu\mu\lambda}^{\rm (cen)}(x_2,x_1,x_3)\; ,
\elabel{4}
\ee
where $\Pi$
is the parity; it is $+1$ if all of the indices of $\gamma$ are ${\rm t}$'s,
or two $\times$'s occur, otherwise it is $-1$. Those compoents of the
shear 3PCF for which $\Pi=+1$ are called even, the others are the odd
components. Note the interchange of the arguments in
(\ref{eq:4}). Furthermore, (\ref{eq:4}) applies equally well to the
3PCF of the shear if measured with respect to a different reference
point attached to the triangle.

We can restate this result in the notation of Takada \& Jain
(2003a,b). Instead of specifying a triangle by its side lengths, one can
also characterize it by two sides and one angle. Let $\psi$ be the angle
at the vertex $\vc X_3$ between the directions to $\vc X_1$ and $\vc
X_2$, where $\psi\in[0,\pi]$ if $\vc x_1\times \vc x_2>0$, and
$\psi\in[\pi,2\pi]$ otherwise. Then, $\gamma_{\mu\nu\lambda}(x_1,x_2,\psi)
=\gamma_{\mu\nu\lambda}(x_1,x_2,x_3)$ for $\psi\in[0,\pi]$, and 
$\gamma_{\mu\nu\lambda}(x_1,x_2,\psi)=\gamma_{\nu\mu\lambda}(x_2,x_1,x_3)$
for $\psi\in[\pi,2\pi]$
in the new notation, and the
parity transformation (\ref{eq:4}) becomes
\be
{\rm P}\eck{\gamma_{\mu\nu\lambda}(x_1,x_2,\psi)} =
\Pi\, \gamma_{\mu\nu\lambda}(x_1,x_2,2\pi-\psi) \;.
\elabel{6}
\ee

\section{E- and B-modes of the shear}
Provided the shear field is due to gravitational lensing by a
(geometrically-thin) matter distribution, it can be derived from a
deflection potential $\psi$, according to
\be
\gamma=\gamma_1+{\rm i}\gamma_2
={1\over 2}\rund{\psi_{,11}-\psi_{,22}}+{\rm i}\psi_{,12}\;,
\elabel{7}
\ee
where indices separated by a comma denote partial derivatives with
respect to the Cartesian coordinates. A shear field with the property
(\ref{eq:7}) is called an E-mode field. A general shear field will not
satisfy (\ref{eq:7}), i.e., will not be derivable from a single scalar
potential.  However, a general shear field can be written as (see
Crittenden et al.\ 2002; Schneider et al.\ 2002; see also Bunn et al.\
2003 for an E/B-mode decomposition of the CMB polarization)
\bea
\gamma&=&\eck{{1\over 2}\rund{\psi^{\rm E}_{,11}-\psi^{\rm E}_{,22}}
-\psi^{\rm B}_{,12}} +{\rm i}\eck{\psi^{\rm E}_{,12}
+{1\over 2}\rund{\psi^{\rm B}_{,11}-\psi^{\rm B}_{,22}}} \nonumber\\
&=& \rund{\gamma_1^{\rm E}+\gamma_1^{\rm B}} +{\rm i}
\rund{\gamma_2^{\rm E}+\gamma_2^{\rm B}} \;,
\elabel{8}
\eea
with the two potentials $\psi^{\rm E,B}$, which are conveniently
combined into a complex field
\be
\psi=\psi^{\rm E}+{\rm i}\psi^{\rm B}\;.
\elabel{9}
\ee
One also defines the Laplacians of the potentials, 
\be
\nabla^2\psi^{\rm E,B}=2\kappa^{\rm E,B}\;;
\elabel{10}
\ee
in lensing, $\kappa^{\rm E}$ is the convergence (or dimensionless
surface mass density) of the deflector. In the case of a matter
distribution extending between sources at high redshift and us, as for
lensing by the large-scale mass distribution in the Universe, the
geometrically-thin lens theory no longer applies; however, both
analytical considerations (e.g., Bernardeau et al.\ 1997; Schneider et
al.\ 1998; van Waerbeke et al.\ 1999), as well as numerical
investigations (e.g., Jain et al.\ 2000) have shown that the cosmic
shear field is very close to an E-mode field; the B-mode
contributions coming from multiple deflections are suppressed by a
large factor compared to the E-mode shear. Furthermore, a small B-mode
contribution derives from the spatial clustering of source galaxies
(Schneider et al.\ 2002), but again is very much smaller than the
E-mode, except at very small angular scales.  However, if the shear is
due not only to lensing, but also to, e.g., intrinsic alignments of
galaxies (e.g., Heavens et al.\ 2000, Crittenden et al.\ 2001; Croft
\& Metzler 2000; Catelan et al.\ 2000), a non-vanishing B-mode can be
expected. Significant B-mode contributions to the cosmic shear signal
have been measured (e.g., van Waerbeke et al.\ 2001a, 2002; Hoekstra et
al.\ 2002; Jarvis et al.\ 2003).

It should be noted that the parity transformation behavior
(\ref{eq:4}) or (\ref{eq:6}) are valid for a general shear field; in
the derivation, no assumption on the E/B-mode character of the shear
field has been made. Also, we have not yet used the fact that the
statistical properties of the shear field are invariant under parity
transformation (i.e., parity-symmetric). 
Takada \& Jain (2003a,b) have argued that the transformation laws 
(\ref{eq:4}), (\ref{eq:6}) only apply for E-modes, and that the signs
are changed for a pure B-mode field. Their argument shall be
reproduced here:

Consider a pure E-mode field, for which (\ref{eq:6}) applies. This can
be turned into a B-mode field by rotating all shears by 45 degrees; in
this operation, $\gamma_{\rm t} \to \gamma_\times$, and $\gamma_\times
\to -\gamma_{\rm t}$. Hence, what was $\gamma_{\rm ttt}$ for the
E-field becomes $\gamma_{\times\times\times}$ for the B-mode field,
implying that $\gamma_{\times\times\times}$ does not change sign under
a parity transformation. This argument, however, is incomplete: the
parity-reversed B-mode shear field is obtained from the parity-reversed
E-mode shear field by rotating the shear by $-45$ degrees, and this
guarantees that (\ref{eq:6}) remains valid even for a pure B-mode
field. As mentioned above, the derivation of (\ref{eq:4}) and
(\ref{eq:6}) makes no assumption about the character of the shear
field, but is derived purely from geometrical considerations (literally,
by drawing triangles with shear sticks attached to the vertices, and
flipping them).

\section{Parity-symmetric shear fields}
In this section we shall consider the consequences of parity-symmetry
for the shear field, in particular with regards to the 3PCF. Consider
first the example just mentioned: take a pure E-mode field, e.g.,
coming from cosmological ray-tracing simulation, and rotate all shears
by 45 degrees, to obtain a pure B-mode field. In the original E-mode
field, peaks and valleys do not occur symmetrically: high-density
peaks are present, but no deep valleys, since the dimensionless
density contrast $\delta$ is bounded from below by $-1$, but clusters
of galaxies have a high density contrast (which, by the way, is the
reason why clusters can be detected by shear measurements, but voids
can not). Hence, the shear field will have tangentially-oriented
patterns around mass concentrations, with no corresponding radial shear
patterns present. The B-field, therefore, will have strong shear
patterns with one circulation, but no corresponding ones with the
opposite circulation. In other words, the B-mode field would carry
circulation information, it would be possible to distinguish between a
right-handed and a left-handed shear field, which obviously violates
parity-symmetry. Hence, the asymmetry between peaks and valleys in the
convergence field $\kappa^{\rm E}$ from which the E-mode shear field
is derived translates into an asymmetry between handedness of the
B-mode field after the 45 degree rotation. If the convergence field
would be symmetric, with as many peaks as valleys, or more precisely,
if all odd moments of $\kappa$ would vanish (as is the case for a
Gaussian random field),
then this handedness
problem in the B-mode field would not occur, and it would be
parity-symmetric. What we shall show in this section is, that all
correlation functions of the form
\be
\ave{\gamma_\mu^{\rm B}(\vc X_1)\dots\gamma_\nu^{\rm B}(\vc X_m)
\gamma_\lambda^{\rm E}(\vc X_{m+1})\dots \gamma_\tau^{\rm E}(\vc
X_{n})}\equiv 0\;,
\elabel{11}
\ee
if $m$ is odd.

\subsection{Parity transformation of E/B-mode shear fields}
Consider a general shear field, which can be a combination of E- and
B-modes. Assume that we flip the field along a line through the
origin, which encloses an angle $\zeta$ with the positive $x_1$-axis. 
A point $\vc x$ is mapped through this flipping to the point $\vc
x'=A\,\vc x$, where the matrix $A$ has the properties that $\det
A=-1$, and $A A=1$ is the unit matrix. This second property shows that
\be
\vc x'=A\,\vc x\; ;\;\;\vc x=A\,\vc x'\;.
\elabel{12}
\ee
Explicitly,
\be
A=\rund{ \begin{array}{cc}
\cos 2\zeta & \sin 2\zeta \\
\sin 2\zeta & -\cos 2\zeta 
\end{array} } \;.
\elabel{13}
\ee
The parity-reversed shear field $\gamma'(\vc x')\equiv ({\rm
P}\gamma)(\vc x')$ can be derived as follows: Let the shear at
position $\vc x$ be $\gamma(\vc x)=|\gamma|{\rm e}^{2{\rm i}\vp}$,
where $\vp$ is the angle the shear `stick' encloses with the positive
$x_1$-axis; the flipped shear is then oriented in the direction $\vp'$
which satisfies $\vp+\vp'=2\zeta$, so that
\be
\gamma'(\vc x')=|\gamma|{\rm e}^{2{\rm i}(2\zeta-\vp)}
={\rm e}^{4{\rm i}\zeta}\,\gamma^*(\vc x)\;.
\elabel{14}
\ee
The same result can of course be directly obtained from
(\ref{eq:transf}), setting $T=A$.

The shear field $\gamma'$ will again be derivable from two fields
$\psi^{E,B\prime}$, for which the relation (\ref{eq:8}) is valid. Hence,
\bea
\gamma_1'(\vc x')&=&{1\over 2}\rund{ {\partial^2 \over \partial x_1'^2}-
{\partial^2 \over \partial x_2'^2}}\psi^{\rm E\prime}(\vc x')
-{\partial^2 \over \partial x_1' \partial x_2'}\psi^{\rm B\prime}(\vc
x')\nonumber \\
\gamma_2'(\vc x')&=&{1\over 2}\rund{ {\partial^2 \over \partial x_1'^2}-
{\partial^2 \over \partial x_2'^2}}\psi^{\rm B\prime}(\vc x')
+{\partial^2 \over \partial x_1' \partial x_2'}\psi^{\rm E\prime}(\vc
x')\,.
\elabel{15}
\eea
The fields $\psi^{E,B\prime}$ will be related to the original ones by
\be
\psi^{\rm E\prime}(\vc x')=\pi^{\rm E}\,\psi^{\rm E}(\vc x)\; ;
\;\;
\psi^{\rm B\prime}(\vc x')=\pi^{\rm B}\,\psi^{\rm B}(\vc x)\; ,
\elabel{16}
\ee
where the factors $\pi^{\rm E,B}=\pm 1$ account for the possibility
that the $\psi$ are not scalar fields, but pseudo-scalars. 
Introducing (\ref{eq:16}) into (\ref{eq:14}) and using
\[
{\partial\over \partial x_i'} \psi(\vc x)
={\partial x_j\over \partial x_i'}\psi_{,j}(\vc x)
=A_{ji}\psi_{,j}(\vc x)\;,
\]
where summation over repeated indices is implied and use has been made
of (\ref{eq:12}), one finds
\bea
\gamma_1'(\vc x')&=&{\pi^{\rm E}\over 2}
\rund{ A_{i1}A_{j1}-A_{i2}A_{j2} } \psi^{\rm E}_{,ij}
-\pi^{\rm B} A_{i1}A_{j2}\psi^{\rm B}_{,ij}\;,\nonumber \\
\gamma_2'(\vc x')&=&{\pi^{\rm B}\over 2}
\rund{ A_{i1}A_{j1}-A_{i2}A_{j2} } \psi^{\rm B}_{,ij}
+\pi^{\rm E} A_{i1}A_{j2}\psi^{\rm E}_{,ij}\;.
\elabel{17}
\eea
Using next the explicit representation (\ref{eq:13}) of the matrix
$A$, one finds that
\[
{\rund{ A_{i1}A_{j1}-A_{i2}A_{j2} }\over 2}\psi_{,ij}
={\psi_{,11}-\psi_{,22}\over 2}\cos 4\zeta + \psi_{,12}\sin 4\zeta\;;
\]
\be
A_{i1}A_{j2}\psi_{,ij}={\psi_{,11}-\psi_{,22}\over 2}\sin 4\zeta 
- \psi_{,12}\cos 4\zeta \;.
\ee
Using these relations in (\ref{eq:17}), one obtains
\bea
\gamma_1'(\vc x')&=&\rund{\pi^{\rm E}{\psi^{\rm E}_{,11}-
\psi^{\rm E}_{,22}\over 2} +\pi^{\rm B}\psi^{\rm B}_{,12} }
\cos 4\zeta \nonumber \\
&+& \rund{ \pi^{\rm E}\psi^{\rm E}_{,12}
-\pi^{\rm B}{\psi^{\rm B}_{,11}-
\psi^{\rm B}_{,22}\over 2} } \sin 4\zeta 
\elabel{19}
\\
&=&
\gamma_1(\vc x)\cos 4\zeta + \gamma_2(\vc x)\sin 4\zeta \;,
\nonumber
\eea
where in the last step we have used the transformation law
(\ref{eq:14}) of the shear. Similarly,
\bea
\gamma_2'(\vc x')&=&\rund{\pi^{\rm B}{\psi^{\rm B}_{,11}-
\psi^{\rm B}_{,22}\over 2} -\pi^{\rm E}\psi^{\rm E}_{,12} }
\cos 4\zeta \nonumber \\
&+& \rund{ \pi^{\rm B}\psi^{\rm B}_{,12}
+\pi^{\rm E}{\psi^{\rm E}_{,11}-
\psi^{\rm E}_{,22}\over 2} } \sin 4\zeta 
\elabel{20}
\\
&=&
\gamma_1(\vc x)\sin 4\zeta - \gamma_2(\vc x)\cos 4\zeta \;.
\nonumber
\eea
Comparing these last two equations with (\ref{eq:8}), one finds that
$\pi^{\rm E}=+1$, $\pi^{\rm B}=-1$: hence, the deflection potential of
the E-mode field is a scalar field, as expected, whereas the potential
of the B-mode field is a pseudo-scalar, so that
\be
\psi^{\rm E\prime}(\vc x')=\psi^{\rm E}(\vc x)\; ;
\;\;
\psi^{\rm B\prime}(\vc x')=-\psi^{\rm B}(\vc x)\; ,
\elabel{21}
\ee
or
\be
\psi'(\vc x')=\psi^*(\vc x)\;.
\ee

\subsection{Consequences for parity-symmetric shear fields}
If a shear field is parity-symmetric, the two fields $\gamma(\vc x)$
and $\gamma'(\vc x)$ have the same statistical properties; in
particular, all their correlation functions are identical.
If one has a pure B-mode shear field, this then implies that all odd
correlation functions vanish: From (\ref{eq:21}) one sees that the
potential of the field $\gamma'$ has the opposite sign of that of
$\gamma$. Since the shear and the potential are linearly related, this
implies that the signs of the two fields $\gamma$ and $\gamma'$ are
opposite, rendering all odd correlation functions zero. More
generally, this argument shows that if we decompose the shear field
into an E- and B-field, as in (\ref{eq:8}), all correlations of the
form
\bea
\ave{\gamma^{\rm B}_\mu(\vc X_1)\gamma^{\rm B}_\nu(\vc X_2)\gamma^{\rm
B}_\lambda(\vc X_3)} &=& 0\; ;\nonumber \\
\ave{\gamma^{\rm B}_\mu(\vc X_1)\gamma^{\rm E}_\nu(\vc X_2)\gamma^{\rm
E}_\lambda(\vc X_3)} &=& 0 \;.
\eea
Therefore, only the pure E-field 3PCF and the mixed correlator
\[
\ave{\gamma^{\rm B}_\mu(\vc X_1)\gamma^{\rm B}_\nu(\vc X_2)\gamma^{\rm
E}_\lambda(\vc X_3)}
\]
are non-zero for parity-symmetric fields. A further generalization
then yields the result (\ref{eq:11}), namely that all correlators
containing an odd number of B-mode shears must vanish.
Furthermore, all odd correlation functions of $\kappa^{\rm B}$ and
$\psi^{\rm B}$ vanish, for the same reason.

One example of this is well known from the 2PCF of the shear; the
mixed correlator $\ave{\gamma_{\rm t}\gamma_\times}$ vanishes
identically due to parity invariance; on the other hand, this
correlation function is linear in $\ave{\kappa^{\rm E}\kappa^{\rm B}}$
(see Schneider et al.\ 2002), and the result derived here applies.

\subsection{Aperture measures}
Given that all third order statistics of the shear are linearly
related to the 3PCF, these results have further implications. Define
the aperture measures for a point at the origin (Schneider 1996) as
\bea
M_{\rm ap}(\theta)&=&\int_0^\theta \d^2\vt\;Q(|\vc\vt|)\,\gamma_{\rm
t}(\vc\vt) \;; \nonumber \\
M_{\times}(\theta)&=&\int_0^\theta \d^2\vt\;Q(|\vc\vt|)\,\gamma_\times
(\vc\vt) \;,
\elabel{24}
\eea
where the tangential and cross components of the shear are taken with
respect tot he direction of the center of the circle of radius
$\theta$, and $Q(\vt)$ is a weight function. These aperture measures
have a number of very useful properties. Employing the relation
(\ref{eq:8}) between shear and the potentials in Fourier space, it is
easy to see that 
\be
M_{\rm ap}(\theta)\equiv 0 \;\hbox{for pure B-field}
\ee
and 
\be
M_{\times}(\theta)\equiv 0 \;\hbox{for pure E-field} \;.
\ee
Therefore, these aperture measures cleanly separate E- and B-modes;
they are therefore used to obtain this separation in data fields, and
essentially all detections of B-modes in second-order cosmic shear
statistics have been made using $\ave{M_{\times}^2}$ (van Waerbeke et
al.\ 2001a, 2002; Hoekstra et al.\ 2002; Jarvis et al.\ 2003). Since
$M_{\rm ap}$ is a scalar quantity, it can be conveniently used to
define higher-order cosmic shear statistics; in Schneider et al.\
(1998), the third-order statistics $\ave{M_{\rm ap}^3}$ was calculated
using quasi-linear perturbation theory of structure growth, whereas
van Waerbeke et al.\ (2001b) calculated $\ave{M_{\rm ap}^3}$ using the
fitting formula of Scoccimarro \& Couchman (2001) for the non-linear
evolution of the bispectrum of the cosmic density fluctuations. Munshi
\& Coles (2003) derived this statistics using the hierarchical
clustering ansatz, which is expected to give an accurate prescription
of the cosmic density field on small scales.

$\ave{M_{\rm ap}^3}$ can be obtained from proper integration over the
E-mode shear 3PCF $\ave{\gamma^{\rm E}_\mu\gamma^{\rm E}_\nu\gamma^{\rm
E}_\lambda}$; similarly, $\ave{M_\times^3}$ can be obtained by
integrating over $\ave{\gamma^{\rm B}_\mu\gamma^{\rm B}_\nu\gamma^{\rm
B}_\lambda}$. The latter, however, vanishes identically for
parity-symmetric shear fields, implying that $\ave{M_\times^3}=0$. For
the same reason, $\ave{M_{\rm ap}^2 M_\times}=0$. For the aperture
measures, the handedness argument used above becomes even more
intuitive. 

Pen et al.\ (2002) employed third-order aperture statistics to the
Virmos-Descart cosmic shear survey; they find a significant non-zero
signal for all their four third-order moments. 
Their measured values of $\ave{M_\times^3}$ and $\ave{M_{\rm
ap}^2M_\times}$ shows that the data is not statistically parity
invariant. Hence, non-zero values of these quantitites 
cannot be accounted for by intrinsic galaxy alignments, as their
correlation functions are expected to be parity-symmetric as well, nor
to higher-order lensing effects -- again, they produce a
parity-symmetric shear field. The remaining explanations are that the
cosmic variance is larger than estimated by Pen et al., or that there
is a yet undetected systematics in the data.

\section{Discussion}
We have shown that for a shear field whose statistical properties are
invariant under parity transformation, all odd correlation functions
of the B-mode shear vanish identically. {This result has been
derived by an explicit calculation of the properties of the
parity-reversed shear field, as well as by a simple consideration:
Whereas an asymmetry between peaks and valleys in the convergence
$\kappa^{\rm E}$, which leads to non-vanishing odd correlation
functions of the E-mode shear field, signifies that matter
overdensities and underdensities behave differently, a similar
asymmetry in the corresponding field $\kappa^{\rm B}$ yields an
asymmetry between the two circularizations of the B-mode shear field,
which for a parity-symmetric field is not permitted.}  In particular,
the 3PCF of the shear vanishes identically for a pure B-mode
field. This result is at variance to some claims in the literature,
and we have attempted to explain the reasons for this.

One possibility to numerically generate a B-mode field which is
parity-symmetric is the following: Take a pure E-mode field, as it is
obtained from ray-tracing simulations, and rotate all shears by 45
degrees. Take a second realization of the same random field (i.e., a
second ray-tracing simulation with identical parameters but different
random number seed), rotate by 45 degrees and parity flip the shear,
or, equivalently, rotate the E-mode field by $-45$ degrees. Then take
the average of these two fields, which is then a pure B-mode field
and parity symmetric (by construction).

The result obtained here provides a very useful diagnostics for
potential systematics in cosmic shear data. Whereas real B-modes can be
present in the data, as they can be generated by intrinsic galaxy
alignments or higher-order lensing effects, e.g. coming from source
clustering, all these effects should obey parity symmetry. A
significant detection of a correlation function involving an odd
number of B-mode components, or a quantity derived from it (such as
the aperture measures discussed above) is a sensitive probe of parity
violation that probably can only be accounted for by systematics. It
should be noted that a similar diagnostics has been used before for
the 2PCF, namely checking that $\ave{\gamma_{\rm t}\gamma_\times}$
vanishes.

\begin{acknowledgements}
I would like to thank Martin Kilbinger, Bhuvnesh Jain, Marco Lombardi,
and Masahiro Takada for very stimulating and helpful discussions.
This work was supported by the German Ministry for Science and
Education (BMBF) through the DLR under the project 50 OR 0106.

\end{acknowledgements}

\end{document}